%
%
\voffset=.3in
\magnification=\magstep1
\def\xx{|x\rangle\langle x'|}
\parskip=4pt plus 2pt minus 2pt 
\centerline{\bf FATHER TIME. I. DOES THE COSMIC MICROWAVE BACKGROUND
RADIATION}
\centerline {\bf PROVIDE A UNIVERSAL ARROW OF TIME ?}
\vskip 2pc
\centerline{T K Rai Dastidar\footnote *{Electronic address : \tt
mstkrd@mahendra.iacs.res.in}}
\vskip 1pc
\centerline{\it Atomic \& Molecular Physics Group, Department of Materials 
Science,}
\centerline{\it Indian Association for the Cultivation of Science,}
\centerline{\it Calcutta 700\thinspace 032, India}
\vskip 3pc
\centerline {\bf Abstract}
\vskip 1pc
It was demonstrated in a recent paper (Mod. Phys. Lett. A{\bf 13}, 1265 
(1998) ; hep-th/9902020) that the existence of a non-thermodynamic arrow 
of time at the atomic level is a fundamental requirement for conservation 
of energy in matter-radiation interaction. In this paper we show that
the Cosmic Microwave Background Radiation can provide this time's arrow
to all particles in the universe that are acted on by the electromagnetic
field.
\vskip 3pc
In quantum field theory, the electromagnetic
field is known to make the Lagrangian density
of, e.g., a complex scalar field covariant under
the U(1) local gauge transformation (GT)
$\exp (-ie\Lambda(x))$, where $x\equiv$ ({\bf r}$,t$). (Note that
a non-dynamical, pure gauge field can do this too [1]).
We showed recently [2] that under a generalisation of U(1) 
to a {\it non-local} gauge transformation\footnote{$^{\rm a}$}{In order to 
remove
possible misunderstandings regarding this notation in [2], we have given
a detailed operational meaning of this non-local GT in Appendix A.}
$$\exp(-ie\Lambda(x))\to\exp (-ie\Lambda\xx)\eqno(1)$$
the covariance property of the Lagrangian is maintained, {\it provided
we replace the ordinary electromagnetic four-potential $A_{\mu}(x)$ by a
non-local potential ${\cal A}_\mu\xx$\/}, which transforms under
the above non-local GT as
$${\cal A}_\mu\xx \to {\cal A}_\mu\xx + \partial_\mu\Lambda\xx +
d_\mu\Lambda\xx \eqno(2)$$
where $\partial_\mu\equiv\partial/\partial x^\mu$, 
$d_\mu\equiv\partial/\partial x'^\mu$. A phenomenological form
of the interaction of matter with this non-local QED
had been given quite a few years ago [3]. The non-local 
electromagnetic field tensor given by
$${\cal F}^{\mu\nu}\equiv{\cal F}^{\mu\nu}\xx 
=(\partial^\mu+d^\mu){\cal A}
^\nu\xx-(\partial^\nu+d^\nu){\cal A}^\mu\xx\eqno(3)$$
{\it remains invariant under the gauge transformation\/} (2)
if and only if
$$[\partial^\mu,d^\nu]=0 \eqno(4)$$
The relation (4) may therefore be considered
as the necessary and sufficient condition for the non-local gauge 
transformation (1) to be a symmetry of the field. (Actually,
if the relation (4) is satisfied, this new symmetry becomes
{\it theoretically self-consistent}~; we are going to address very shortly
the question whether {\it this symmetry is, in fact, obeyed by nature}.) 
Before proceeding further, let us recall the three important results 
obtained in [2]~:
\item {i)} As a consequence of matter-(non-local radiation) interaction,
a correlated two-photon absorption {\it linear in intensity} occurs.
(This had already been predicted in [3].) Direct experimental 
evidence for such two-photon absorption exists [4]. 
\item {ii)} The {\it principle of causality\/} emerges as a necessary 
condition 
for energy conservation in such correlated two-photon absorption processes~; 
out of the entire conceivable range of non-local temporal 
correlations (which spans both the past and the future), only
the past ensures energy conservation in matter-radiation interaction 
events of the above kind, thus defining a non-thermodynamic arrow of 
time in the quantum (atomic) level. 
\item {iii)} The self-consistency relation
(4), which implies that the two space-time points $x,x'$ are
completely independent (i.e. need not lie within the light cone of one
another), shows that the non-local correlation must be of the
Einstein-Podolsky-Rosen (EPR)-type. It was shown in [2] that
this EPR-like character of the non-local
correlation in the new QED {\it is another necessary condition for energy
conservation\/}. 

Note that the results (i) and (iii) address the question mentioned after 
equation (4). The result (iii) is very closely linked with the
result (ii), and the present paper hinges on the arrow of time given
in the latter~; for completeness we re-derive the results (ii) and (iii)
in the appendix B.

The purpose of the present paper is to show that the cosmic microwave 
background radiation (CMBR) provides a {\it via media} for imparting 
each and every particle (which is acted on by an electromagnetic field)
in the universe with this arrow of time. We proceed to
demonstrate this by recalling from a recent work [5] that the 
interaction of matter 
with CMBR fits into the nonlocal picture as laid out in [2], and thus
allows for correlated two-photon absorption by all particles, 
whereupon the time's arrow is established by energy conservation.

The substance of [2] is that if a radiation field (either as generated, or as 
detected by atomic electrons) enjoys a large second-order coherence
--- in other words, two photons far apart in space can be {\it coherently} 
absorbed by the atomic electrons within a phase 
difference $\Delta\phi=\omega\delta t \ll 1$ ($\delta t$ being the time
interval between the two photon-absorption events) --- then the field 
is describable by the 
nonlocal QED. In practice, this translates to the condition that
the photon density be significantly higher than the absorber (electron)
density. (Of course, if the effect is to be measurable,
the electron density must at the same time be high enough to
make the observation of the two-photon absorption process experimentally
viable.) Some typical matrix
elements for such correlated two-photon absorption by atoms have been 
worked out in [2]. The question is~: under what
circumstances can this coherence condition be satisfied in practice ?
Two suitable laboratory settings have already been described in
[2], namely, (i) a laser field so intense that the average 
photon density is much larger than the electron density, and (ii) a
squeezed light field with phase-correlated photon pairs, e.g. {\it signal} 
and {\it idler} photons. Very recently we have shown [5] that a 
third, {\it cosmological} setting is provided by this universe through the
{\it via media} of the cosmic microwave background radiation (CMBR).
Indeed, we can call the CMBR the ``Lowest common denominator'' from the 
viewpoint of this paper, in the sense that it influences {\it each
particle of the universe capable of interacting with an 
electromagnetic field\/}. Note that, in the entire universe, the
overall photon density is several orders of magnitude higher than
the baryon or lepton density. As such, if we remember that the CMBR 
{\it is incident from all directions\/}, it is obvious that so long as 
the matter density is tenuous enough (so that the photon density remains 
higher than the matter density), two distinct photon 
absorption events can always take place at two arbitrary space points 
so that the above second-order coherence condition is satisfied.
For example, it is easy to 
see that if two millimetre-wave photons are absorbed within a gap of 
$\sim 0.1-1$ ps, then the phase gap between them is $\ll 1$ [6].
The difference between the laboratory settings and the cosmological
setting is that, whereas in the former the spatial distance between 
the two distinct (photon absorption) events is of the order of
atomic/molecular dimensions, in the latter the said distance can be of
macroscopic dimensions. 

In [5] the consequences of such a correlated
two-photon absorption loss upon 
the continuous spectrum of a blackbody radiation travelling
through gaseous matter were worked out. It was shown there that the
blackbody radiation energy density $E(\nu)$ at a frequency $\nu$ which, 
in the absence of any absorbing matter, is given by the Planck formula, 
would now be reduced to 
$$G(\nu)=E(\nu)-F(\nu),\quad F(\nu)=\alpha E(\nu /2) \eqno(5)$$
where $\alpha$ is an effective correlated
two-photon absorption factor for the intervening medium. For the
present we need not go into the detailed dynamics needed for calculating
$\alpha$ ; we only note that it will depend, among other things, on the
{\it density of the absorber\/}. Detailed expressions for $E(\nu),
F(\nu)$ and $G(\nu)$ have been given in [5]. It was shown that the theory
predicts an apparent shift in the ``blackbody temperature'' of the 
radiation travelling through the gas, and an experimental test for
checking this prediction was suggested.

Is there any observational evidence that such a two-photon absorption 
from the CMBR really does take place anywhere in the universe ? There is, 
as has been demonstrated in [5]. To recapitulate briefly, such a two-photon 
absorption loss superimposed upon the spectrum of the CMBR travelling
through interstellar and intergalactic matter
manifests itself as an apparent deviation  of the 
spectrum from the Planck formula --- the magnitude of the deviation
depends, of course, on the value of the absorption factor $\alpha$ --- 
and hence this apparent deviation should be more ``visible''
when the CMBR is observed from within some dense gaseous atmosphere 
than from within more
tenuous atmospheres. We now draw the reader's attention
to the balloon-borne measurements of the cosmic microwave spectrum 
carried out by Woody and Richards [7] and the measurements
carried out in the {\it COBE\/} (Cosmic Background Explorer) satellite
[8], which fit into these two categories admirably. 
The spectrum observed by Woody and Richards [7] (see Fig. 2 
therein) shows a distinct dip
below the Planck curve over extended regions towards the right of the
spectral peak, the maximum deviation occuring around {\it twice the peak 
frequency\/} --- exactly as implied by the eqn.\thinspace (5) --- 
whereas in the {\it COBE\/} measurements the deviations 
from the Planck curve are much more miniscule. 

To avoid misgivings that we are jumping to conclusions, 
we emphasize two points. First, whether such a deviation from the 
Planck distribution as predicted by
our theory does occur at all is a question that can only be settled by
experiment~; one possible experimental test has been suggested in [5].
Secondly, the {\it viability\/} of such a correlated two-photon 
absorption process under conditions of Woody and Richards' balloon 
measurements has also been established in [5]. Therefore, subject to the
condition that our prediction is experimentally verified, we can claim
that the CMBR travelling through tenuous matter can satisfy the condition
for second-order coherence and hence, as laid down by the non-local 
QED, is capable of giving rise to {\it correlated two-photon
absorption throughout the universe} --- thus
offering a very simple picture of the arrow of time in the entire 
universe at the quantum (atomic) level {\it as a necessary condition
for energy conservation\/}. 

Our picture does not depend on any particular cosmological model.
However, if one couples it to the Big Bang model, it would appear
that the time's arrow (and hence {\it causality\/}) were set in
motion as soon as the universe started off --- the latter ``creating'', 
as it were, its own flow of time as well as its own physical laws as 
soon as it was born. Appealing as the picture is, it is still incomplete ; 
until the non-local gauge symmetry of the universe can be extended 
beyond eqn.\thinspace (2) to cover the 
entire range of electroweak interactions, our theory remains confined 
only to those constituents of the universe that are acted on by the 
electromagnetic field and hence cannot embrace neutrinos, which 
are generally supposed to make up the {\it cold dark matter\/} in the
universe. Work is in progress towards this goal.
\vskip 1pc
The author is indebted to Dr Krishna Rai Dastidar, Department of
Spectroscopy for useful discussions and to Dr J Chakraborti, Department
of Theoretical Physics for his interest. A correspondence with Prof.
Alex Dalgarno has helped to clarify certain ideas.
\vfill\eject
\centerline{\bf Appendix A}
\vskip 1pc
In the earlier papers (MPLA {\bf 13}, 1265 (1998) and {\bf 14}, 1193 (1999))
we had introduced the non-local Gauge Transformation operator $T=\exp   
(-ie\Lambda\xx)$ simply as a shorthand notation for an {\it integral operator}.
To avoid any possible misunderstandings arising out of our notation,
we further clarify below the operational meaning of such non-local integral
operators acting on an arbitrary field $\phi(x)$. First of all, 
$|x\rangle,\langle x'|$ are {\it not} Dirac ket and bra vectors~; and
in general, the operational meaning of any non-local (integral) operators
$P\xx$, $Q\xx$ is given by the expressions
$$P\phi\equiv\int P(x,x')\phi(x')\thinspace d^4x',\qquad PQ\phi\equiv
\int\int P(x,x')Q(x',x'')\phi(x'')\thinspace d^4x'\thinspace d^4x''\eqno(A1)$$
Thus the effect of the non-local gauge transformation $T$ on $\phi$ is
given by
$$T\phi\equiv\int e^{-ie\Lambda(x,x')}\phi(x')\thinspace d^4x'$$
where $\Lambda(x,x')$ can be any arbitrary scalar function of $x$ and $x'$.
Thus, for infinitesimal $\Lambda$, we have $\delta\phi=T\phi-\phi=
-ie\int\Lambda(x,x')\phi(x')\thinspace d^4x'$.
One at once notes the parallelism with the Lippmann-Schwinger equation
in the first order~:
$$\eqalignno {\Psi&=(1+G_0V)\psi\cr &=\psi(\vec r)+\int G_0(\vec r,\vec r')
V(\vec r')\psi(\vec r')\thinspace d^3r'\cr}$$
where $\psi$ is an unperturbed state, $\Psi$ is a scattering state,
$G_0(\vec r,\vec r')$ is the non-local Green's function $(E-H_0)^{-1}$
and $V$ is the scattering potential.

Similarly, the newly defined derivative 
$$\eqalignno{D_\mu\phi &=\partial_\mu\phi(x)+d_\mu\phi(x)+
ie{\cal A}_\mu\xx\phi\cr
&=\partial_\mu\phi(x)+d_\mu\phi(x)+
ie\int{\cal A}_\mu(x,x')\phi(x')\thinspace d^4x'\cr}$$
($d_\mu=\partial/\partial x'^\mu$, as defined in the Appendix A of the
earlier paper) is covariant, provided ${\cal A}$ transforms under this
gauge transformation as
$${\cal A}_\mu\phi\to{\cal A}_\mu\phi+\int[\partial_\mu\Lambda(x,x')]\phi(x')
\thinspace d^4x'+\int[d_\mu\Lambda(x,x')\phi(x')]\thinspace d^4x'$$
as may be easily checked using eqn.\thinspace (A1).  (Of course, for
local $\phi$, the second term on the right side of $D_\mu\phi$ vanishes.)
All the results derived in the earlier papers and the present paper
are obtained without assuming any specific functional form of the quantities
$\Lambda(x,x')$ and ${\cal A}(x,x')$. 
\vfill\eject
\centerline{\bf Appendix B}
\vskip 1pc
We briefly recapitulate here, following [2], how the time's
arrow and the EPR-like character of the non-local correlation in the
new QED appear in the theory as necessary conditions for conservation of 
energy in the atomic scale.
A matrix element for matter-radiation interaction
with the electromagnetic field potential ${\cal A}(x,x')$ (which
transforms as eqn.\thinspace (2)) is typically of the form
$$M_{fi}(t)\propto\langle\psi_f(x),n_f|{\hat p}.
{\cal A}(x,x')|\psi_i(x'),n_i\rangle,$$
the integration running over $d^3r$ and $d^4x'$. 
Special relativity requires that the time integral over $dt'$ should run 
from $-\infty$ to $(t-\rho/c)$ [where $\rho=|${\bf r} -- {\bf r}$'|$] 
for the retarded interaction, and from
$(t+\rho/c)$ to $+\infty$ for the advanced interaction. After Fourier
expanding ${\cal A}$ in two sets of photon modes
(see, e.g. [9]), the
time integration over $dt$ gives us the energy-conserving
delta-function $\delta(E_f,E_i+\hbar\omega+\hbar\omega')$ in the 
two-photon absorption process (see below).
If we agree to define that the atom has been excited,
i.e. the matrix element has come into existence, at the time $t=0$,
then it is at once obvious that the the time integral 
over $dt$  can only run from 0 to $\infty$, giving finally a
double integral to be chosen from
$$\int_0^\infty f(t)dt \left (\int_{-\infty}^{t-\rho/c}g(t')dt',\quad
\int_{t+\rho/c}^\infty g(t')dt'\right )$$
where $f(t)=\exp\left [{i\over \hbar}(E_ft-\hbar\omega t)\right ], 
g(t')=\exp\left [-{i\over \hbar}(E_it'+\hbar\omega't')\right ]$.
It is obvious that the retarded interaction gives the
required $\delta$-function for energy conservation, while the
advanced interaction fails to do so. Thus we are constrained to limit 
the non-local temporal correlation {\it to the past only\/}~; 
the fundamental principle of energy conservation
has given us an arrow of time, i.e. the principle of causality, as
a corollary.\par
However, one must not lose sight
of a very important point~: {\it x} and {\it x}$'$ need not be
signal-linked. This is contrary to the spirit of special relativity, 
and requires an explanation. We note that if {\it x} and {\it x}$'$ 
are constrained to be so linked, then for $\rho > ct$ (remember that {\bf r} 
and {\bf r}$'$ extend over all space)  
the upper limit of the retarded $t'$-integral becomes negative and
the integral vanishes, leading to non-conservation of energy~!
We are therefore obliged to re-write the retarded integral as
$\int_{-\infty}^tg(t')dt'$. 
This is the {\it physical reason\/} why the non-local
correlation must be of the ``spooky'' EPR-type~; the fact that this
characteristic is required for {\it theoretical self-consistency\/}
has already been mentioned in connection with eqn.\thinspace(4).
\vfill\eject
\centerline {\bf References}
\item{1.} T K Rai Dastidar and Krishna Rai Dastidar, Mod. Phys. 
Lett. A{\bf 10}, 1843 (1995) ; Nuovo Cimento {\bf 109B}, 1115 (1994)
\item{2.} T K Rai Dastidar and Krishna Rai Dastidar, Mod. Phys. 
Lett. A{\bf 13}, 1265 (1998) ; {\it Errata} A{\bf 13}, 2247 (1998) ;
hep-th/9902020
\item{3.} (a) T K Rai Dastidar and Krishna Rai Dastidar, 
Abstracts of XI Int. Conf. At. Phys. (Paris, 1988), p. I-23 ; 
(b) Krishna Rai Dastidar {\it in} ``Advances in Atomic and Molecular 
Physics'', ed. M S Z Chaghtai (Today's and Tomorrow's Publishers, 
New Delhi, 1992) p. 49.
\item{4.} N P Georgiades, E S Polzik, K Edamatsu, H J Kimble and
A S Parkins, Phys. Rev. Lett. {\bf 75}, 3426 (1995)
\item{5.} T K Rai Dastidar, Mod.\thinspace Phys.\thinspace Lett. A
{\bf 14}, 1193 (1999)~; quant-ph/9903043
\item{6.} Remember that there need not be any causal link 
between the two events --- indeed, in most cases there {\it cannot be}.
Two events at two space-time points 
{\it x},\thinspace{\it x}$'$ can be phase-matched without 
any causal correlation, i.e. without the events' ``knowing'' about each 
other. As a matter of fact, this furnishes the clue to the EPR-like 
character of the nonlocal QED.
\item{7.} D P Woody and P L Richards, Phys. Rev. Lett. {\bf 42}, 925
(1979)
\item{8.} J C Mather et al, Astrophys. J. {\bf 420}, 439 (1994)
\item{9.} I M Gel'fand and G E Shilov, {\it Generalised Functions}
(Acad. Press, 1964) Vol. 1, Chap. 2, eq. 1.3(1)
\end